# Principal Components and Independent Component Analysis of Solar and Space Data

A.C. Cadavid[1], J.K. Lawrence[1], and A. Ruzmaikin[2]

[1]*Department of Physics and Astronomy, California State University, Northridge, 18111 Nordhoff Street, Northridge, California 91330-8268, U S A (e-mail: ana.cadavid@csun.edu)*

[2]*Jet Propulsion Laboratory, California Institute of Technology, 4800 Oak Grove Drive, Pasadena, California 91109, USA*

**Abstract.** Principal Components Analysis (PCA) and Independent Component Analysis (ICA) are used to identify global patterns in solar and space data. PCA seeks orthogonal modes of the two-point correlation matrix constructed from a data set. It permits the identification of structures that remain coherent and correlated or which recur throughout a time series. ICA seeks for maximally independent modes and takes into account all order correlations of the data. We apply PCA to the interplanetary-magnetic-field polarity near one AU and to the 3.25 $R_\odot$ source-surface fields in the solar corona. The rotations of the two-sector structures of these systems vary together to high accuracy during the active interval of solar cycle 23. We then use PCA and ICA to hunt for preferred longitudes in northern hemisphere, Carrington maps of magnetic fields.

## 1. Introduction

Principal components analysis (PCA) and independent component analysis (ICA) seek to identify global patterns in sets of "images," whether these are spatial images, such as magnetograms, or segments of time series as in solar-wind data.

      PCA searches for orthogonal modes of the two-point correlation matrix constructed from a data set and permits the identification of structures that remain coherent and linearly correlated or which recur throughout a time series. These modes, or empirical orthogonal functions (EOFs), are ordered according to the degree of linear correlation with the whole data set. The first benefit of the procedure is that it may allow the summary or capture of the essential nature of the data set using a significantly reduced subset. The second benefit is that the procedure will lead to EOFs that are amenable to physical interpretation, that is, that it will extract some hidden meaning from the data. For example, the EOFs may be eigenmodes of the underlying physical process. This second form of success is rarer, but there are techniques to help it along. In particular, the structures obtained from PCA in general are not independent because of possible

higher order correlations. ICA addresses this point by seeking maximally independent modes, taking into account all order correlations of the data. This approach re-combines a reduced set of EOFs to produce a new set of maximally non-Gaussian, or maximally independent, images called independent components (ICs).

Previously, we have applied PCA to the large-scale, axisymmetric magnetic field and related this to the solar dynamo (Lawrence, Cadavid, and Ruzmaikin, 2004). Subsequently, we applied ICA to these fields and found common periodicities between their time variations and the characteristic periodicities in the solar wind and interplanetary magnetic field (Cadavid, *et al.*, 2005a,b). In the present paper we will provide details of these methods and apply them to new problems.

In Section 2 below we will apply PCA to two different, but physically related, data sets: the interplanetary magnetic field (IMF) polarity at one AU, and the 3.25 $R_\odot$ source-surface fields from magnetic images of the Sun. We will show that similar PCA modes can be found in the two data sets and that they can be related to demonstrate a close physical connection. In Section 3 we shall apply PCA and ICA to magnetic activity in a particular latitude band in the Sun's northern hemisphere in a search for the presence of active longitudes. The EOFs are not easily interpretable, but the ICs focus on specific longitudes. A close look at the time variations of the importance of these ICs in the data shows some periodic recurrence of activity lasting from two to five hours. Our results are summarized and conclusions are drawn in Section 4.

**2. Principal Components Analysis**

2.1 Mathematical Background

The purpose of PCA is to reduce the original data set of two or more sequentially observed variables by identifying a small number of "meaningful" components or modes (*e.g.* Jackson 2003). The method therefore permits the identification of coherent structures as dominant or recurring patterns in a time series. To start, the data is organized in an $m \times n$ matrix $X^T(x,t) = [u_1 \cdots u_n]$, where the $u_i(x)$ vectors describe the $m$ pixels in an image observed at $n$ different times. To be precise in this discussion, and without loss of generality, we can assume that the data matrix has dimension $m \geq n$ with rank $r \leq n$. The data is then centered by subtracting the average from each image. The formal procedure consists in solving the eigenvalue problem for the two-point $m \times m$ covariance matrix $C(x,x') = \sum_t X^T(x,t)X(x',t)$. In general the number of eigenvalues will be the same as the rank $r \leq n$ of the data matrix. When dealing with linearly-independent observations we obtain $r = n$ eigenvalues $\lambda_l$. Otherwise the number of eigenvalues is equal to the rank of the data matrix. The corresponding $m$-dimensional eigenvectors $\phi_l(x)$ are identified as the spatially-dependent "empirical orthogonal functions" (EOFs). The alternative $n \times n$ covariance matrix $C'(t,t') = \sum_x X(x,t)X^T(x,t')$ leads to the same eigenvalues with corresponding $n$-dimensional orthogonal vectors $e_l(t)$. These are identified as the principal components (PCs) which describe the time dependence of each mode. The data can then be decomposed in terms of the modes as:

$$X^T(x,t) = \sum_{l=1}^{n} \sqrt{\lambda_l} e_l^T(t)\phi_l(x) \qquad (1)$$

If the power of the sources in the data is stronger than that of the noise, the sequence (spectrum), of ordered eigenvalues will exhibit a clear drop or "breakpoint" at a particular mode. The dominant EOFs associated with the large eigenvalues before the breakpoint will mainly characterize the data and will carry a large percentage of the variance. The remaining EOFs will contain transient fluctuations or noise. Upon identification of the leading modes it is possible to describe the system in terms of this lower number of degrees of freedom or coherent structures.

The $n$ EOFs and PCs can be organized in a $m \times n$ matrix $\Phi(x) = [\phi_1 \cdots \phi_n]$ and an $n \times n$ matrix $E(t) = [e_1 \cdots e_n]$ respectively, while $\Lambda$ is the $n \times n$ diagonal matrix of eigenvalues and $L$ is an $n \times n$ diagonal matrix with $L_l = \sqrt{\lambda_l}$ in its entries. In the general case $L_l > 0$ for $1 \leq l \leq r$ and $L_l = 0$ for $(r+1) < l \leq n$. With these definitions, and working with normalized eigenvectors, the following relations are satisfied:

$$C\Phi = \Phi\Lambda$$
$$C'E = E\Lambda \qquad (2)$$

and

$$X^T = \Phi L E^T \qquad (3)$$

Working with centered data and normalized eigenvectors we can identify the last equation with the singular value decomposition (SVD) of the data matrix (Elsner and Tsonis, 1996). In this context, the columns of $\Phi$ are called the left singular vectors, and form an orthonormal basis. The rows of $E^T$ are the principal components which are now identified as the right singular vectors.

2.2 Application of PCA to Heliospheric and Solar Data

*2.2.1 Solar Wind / IMF Longitude*

As an illustration of the techniques, we will apply them to two related data sets and then use them to make a comparison. The first data set that we consider is $B_L$, the longitude of the average interplanetary magnetic field (IMF). We use the hourly averaged data (longitude angle) from the *Wind* and ACE spacecraft near one AU (from the OMNI2 website http://omniweb.gsfc.nasa.gov/). In Figure 1 we show these hourly data in rasters of $m = 655$ pixels, corresponding to 27.2917 days. They differ in duration by 23.616 minutes from the solar synodic Carrington period (27.2753 days). The data are highly bimodal; the light (dark) colored points in Figure 1 indicate that the field points inward toward (outward from) the Sun.

In carrying out a principal components analysis of these data, we regard each of the $n = 131$ individual 655-hour rasters as a separate one-dimensional image of the IMF polarity near one AU. The PCA yields 131 eigenvalues; the first 20 are shown in Figure 2.

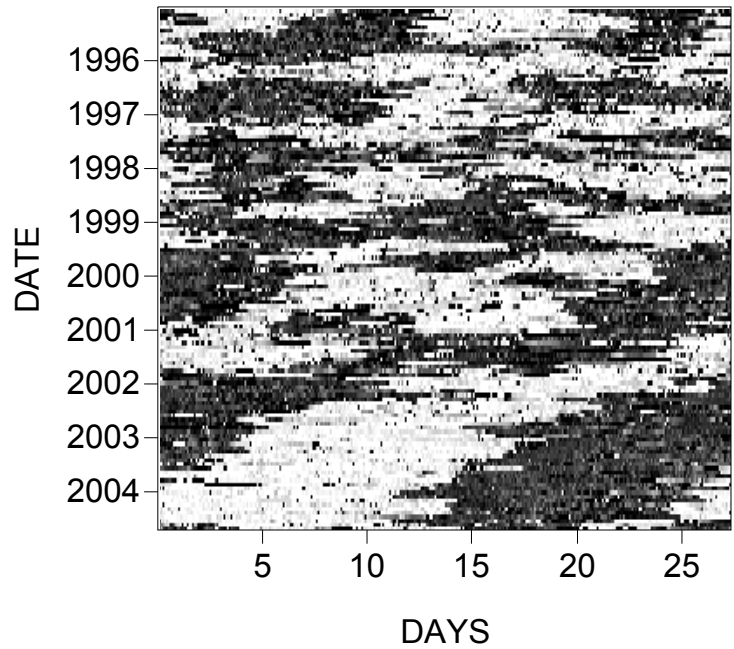

*Figure 1. Interplanetary magnetic field polarity, in Geocentric Solar Ecliptic coordinates, at the Wind and ACE spacecraft. Each of the 131 rasters represents 655 hours or 27.292 days, roughly a Carrington period. Light (dark) pixels indicate that the field points inward toward (outward from) the Sun.*

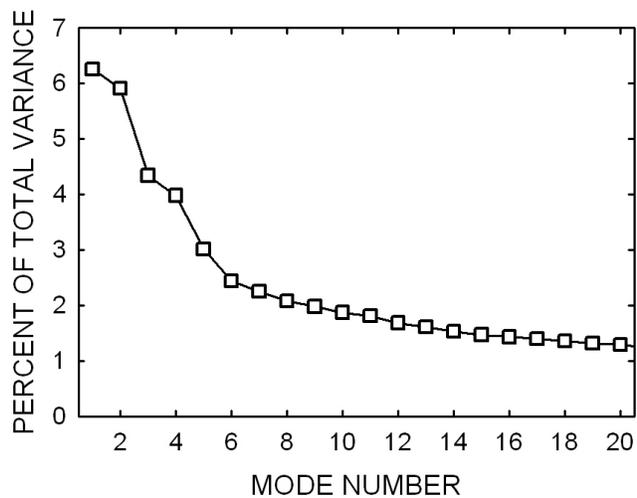

*Figure 2. The leading 20 eigenvalues in the spectrum obtained in the PCA of the interplanetary magnetic field. The eigenvalues give the percent of total variance accounted for by each mode.*

The leading two eigenvalues, each accounting for about six percent of the total variance of the data, will be seen to represent a related pair. The EOFs corresponding to these modes are shown in Figure 3.

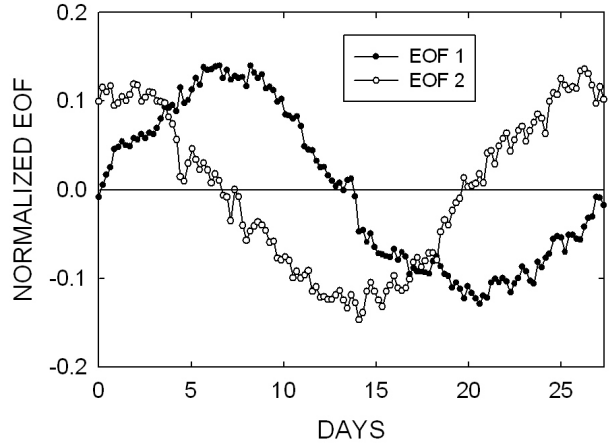

*Figure 3. The first two empirical orthogonal functions obtained in the PCA of the IMF polarity near 1 AU. These sine-wave structures with quarter-wave shift indicate a precessing, two-sector mode.*

These two EOFs consist of a pair of single-period sine waves with a 90° phase shift between them. The corresponding PCs (Figure 4) give the time dependence of the strength of each of these EOFs in the data.

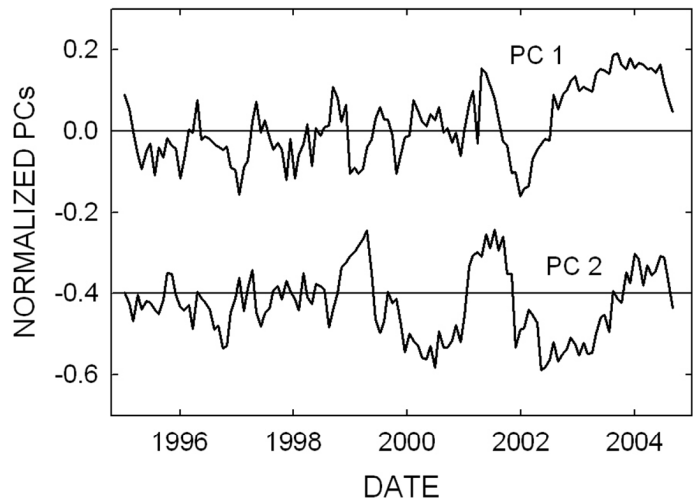

*Figure 4. The first two principal components from the PCA of the interplanetary magnetic field. For clarity in the presentation PC2 has been shifted vertically.*

The relative phasing of these two PCs indicates that the two modes, taken together, represent a two-sector IMF structure that rotates at other than the Carrington rate.

The modes corresponding to eigenvalues 3 and 4 in Figure 2 also form a related pair. The corresponding EOFs consist of two two-cycle sine waves, and these correspond to a rotating four-sector structure in the solar wind (SW)/IMF. The phase lag of the two leading modes can be estimated by associating PC1 and PC2 with sine and cosine time series and finding the phase as an arctangent. When a jump of more than $\pm\pi$ radians was encountered, we, somewhat arbitrarily, added or subtracted $2\pi$ to make the jump less than $\pi$ in magnitude. The phase drift of the IMF longitude from 1997 to 2004 is plotted as the solid circles in Figure 7.

*2.2.2 Wilcox Solar Observatory Coronal $3.25R_0$ Magnetic Carrington Maps*

The second data set we will investigate here, in order to compare to the preceding, is the set of Wilcox Solar Observatory Carrington maps (from http://wso.stanford.edu) projected from the photosphere to a source surface at $R_S = 3.25\ R_\odot$. The fields on this source surface indicate open magnetic fields extending into the heliosphere. They represent a source for the quasi-static background IMF on which more dynamical and irregular phenomena such as magnetic clouds occur. We have included $n = 396$ Carrington Rotations (CR) 1642 – 2037. Each map comprises 30 rows of 72 pixels. Rasters 1 – 15 span from the north pole to the Equator of the Sun. Rasters 16 – 30 span from the equator to the south pole. Each vertical row represents a range of sine latitude of $1/15 = 0.066$. Each horizontal pixel represents 5° of longitude or 9.092 hours of time.

The PCA of the two-dimensional images is handled by unwinding the image into a $m = 30 \times 72 = 2160$ pixel one-dimensional string and proceeding as in the previous section. The leading eigenvalue accounts for 38% of the total variance in the data set. Eigenvalues 2 and 3, accounting for 11.5% and 9.9% of the variance appear to form a related pair, similar to the IMF data discussed before. Eigenvalue 4 accounts for 5.1% of the variance, and the rest decrease from there. The leading three EOFs are shown in Figure 5.

The leading EOF 1 in Figure 5 corresponds to a magnetic dipole oriented North – South. It is very close to a spherical harmonic $Y_n^m(\theta,\phi)$ with $n = 1$ and $m = 0$. EOFs 2 and 3 are equivalent to dipoles in the equatorial plane with a quarter wave shift. They are roughly combinations of the spherical harmonics with $n = 1$ and $m = \pm 1$. There is no $n = 0$ (monopole) contribution. It is to be expected that the $n = 1$ harmonics will be the most prominent at $r = 3.25\ R_\odot$, due to "coronal filtering" (Wang, et *al.*, 1988). This arises because the corona behaves in such a way that the fields therein are approximately potential fields. If the potential is expanded in spherical harmonics, then solving the Laplace equation in the space external to the photosphere gives solutions that are sums of terms proportional to $Y_n^m(\theta,\phi)r^{-(n+1)}$. Thus, the greater the index *n* the faster the contribution dies with distance (Schatten, Wilcox, and Ness, 1969; Altschuler and Newkirk, 1969).

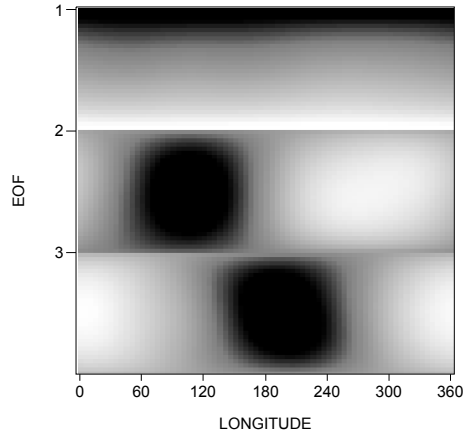

*Figure 5. Three leading EOFs of the $3.25R_0$ coronal field synoptic maps from WSO. EOF1 corresponds to the N - S polar dipole ($n = 1, m = 0$). EOFs 2 and 3 (including $n = 1, |m| = 1$) have the form of dipoles lying in the equatorial plane. Note that EOFs have arbitrary sign.*

Figure 6 shows the PC time series for EOFs 1, 2, and 3 for the span 1976 to 2005. We see that mode 1 just represents the polar dipole field and indicates its 20-year cycle. It has reversed polarity near the beginning of 1980, 1990, and 2000. The PC2 and PC3 time series, in Figure 6, show that modes 2 and 3 taken together represent a magnetic dipole in the solar equatorial plane that is rotating about the polar axis at variable rates not always matching the Carrington rate. Using these we estimate the phase lag just as we did for the IMF polarity data, and the result is shown in Figure 7 as the open circles. In Figure 7 a horizontal plot represents rotation at the Carrington rate. The negative slope indicates a shorter period, near 27.0 days.

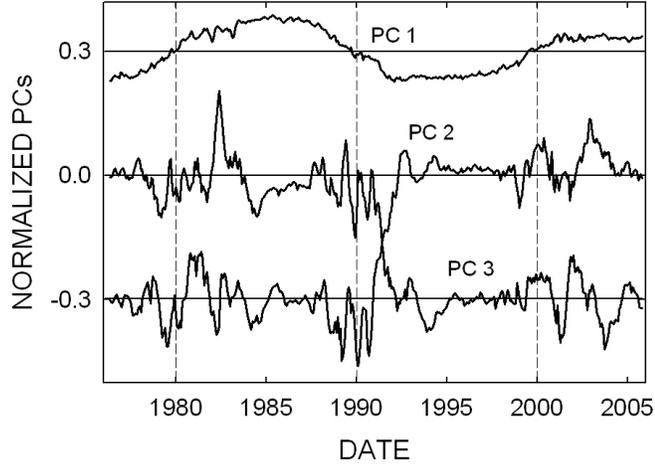

*Figure 6. Time series for the PCs 1, 2, and 3 of the coronal field synoptic maps from WSO. For clarity in the presentation PC1 and PC3 have been shifted vertically.*

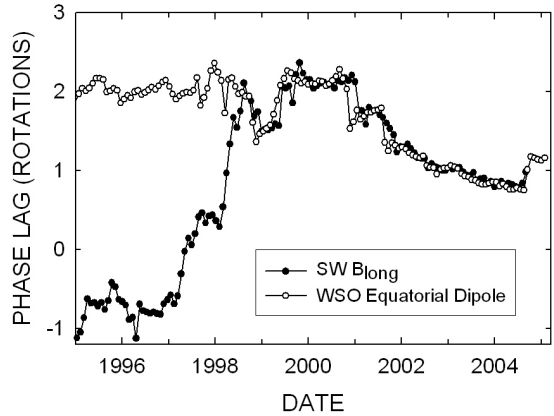

*Figure 7. Phase lag versus time for PC1 and PC2 of the IMF (solid circles) and for PC2 and PC3 of the potential field extrapolated, coronal magnetic field (open circles).*

The phase lags for the IMF and WSO data (Figure 7) were shifted to bring the two plots together during the years 1998 – 2004. We can see that from mid-1998 to at least mid-2004 the two phases of the coronal source-surface equatorial dipole and the IMF two-sector structure track each other extremely closely and in considerable detail. In the period before 1998, the two plots do not track at all. It is no doubt relevant that during the quiet-Sun period between activity cycles 22 and 23 the amplitude of the source surface dipole was relatively weak $A \equiv (PC2^2 + PC3^2)^{1/2} \approx 0.5$. This amplitude increased beginning in 1998.2 to an erratic $A \approx 3$.

## 3. Independent Component Analysis

3.1 Mathematical Background

PCA, based on the linear correlation between pairs of data points, offers a way to extract structures that remain spatially coherent or recur many times throughout a time series. Although required to be linearly uncorrelated, due to higher order correlations these structures are not necessarily independent.

ICA (Hyvärinen, Karhunen, and Oja, 2001; Stone, 2004 and references therein) is based on the assumption that source signals due to different physical processes are statistically independent in the sense that the value of one signal gives no information about the values of the others. Mathematically, two variables $x$ and $y$ are statistically independent if their joint probability density function (pdf) is the product of their individual pdfs:

$$p(x, y) = p(x)p(y) \qquad (4)$$

This implies that the mean (expectation value) of any moment of the two variables

$$E[x^p y^q] = \iint p(x, y) x^p y^q \mathrm{d}x\mathrm{d}y \qquad (5)$$

can be given by the product of the individual expectation values

$$E[x^p y^q] = \iint p(x)p(y) x^p y^q \mathrm{d}x\mathrm{d}y = E[x^p]E[y^q] \qquad (6)$$

If the variables are not independent, but are merely uncorrelated, this equality is valid only for $p = q = 1$, that is $E[xy] = E[x]E[y]$. The Gaussian, or normal, distribution is entirely determined by its first two moments (mean value and standard deviation). It follows from this that uncorrelated Gaussian variables are independent. For non-Gaussian variables, all moments may be needed to specify the pdfs, and higher-order correlations must be taken into account to establish independence.

Actual observations of a system are composed of mixtures of source signals. ICA aims to extract those statistically-independent signals from the observed mixtures and thus to get information about the underlying physical processes. To achieve this, ICA makes use of the Central Limit Theorem. Given two or more signals with different pdfs, the pdf of the sum of these signals tends to be more Gaussian than those of the constituent signals. ICA is a method to extract from the more Gaussian signal mixtures, the combinations with the most non-Gaussian possible pdfs. These are identified as the "independent components" (ICs) of the observations. ICA has been applied to varied problems, such as recovering individual voices from mixtures recorded by arbitrarily-located microphones in a crowded room, or centers of brain activity from brain waves recorded by multiple sensors on the scalp.

As will be shown later, we will take the magnetic activity in each of 71 solar-longitude bins as our variables and the values of these observed in 406 successive solar rotations as the 71 observed signals. Each of these 71 signals will have a pdf, which is calculated from a histogram of the 406 observed values of that signal. The goal of the ICA is to recombine the 71 observed signals into a set of global modes (the ICs), such that the fluctuations of the modes (the "mixing vectors," or MVs) have maximally non-Gaussian pdfs. The most non-Gaussian of these may be taken to represent source signals.

Mathematically, $X(x,t) = [u_1 \cdots u_n]^T$ is the observed data, which can be described by the relation $X = AS$, where $S$ is a matrix containing the unknown source signals and $A$ is an unknown mixing matrix that mixes the sources to give the observations. The problem is finding the "unmixing matrix" ($W$) such that $S = WX$. Each source signal can then be obtained by taking the inner product of an "unmixing vector," a row in $W$, with the mixed signals in the data. To gain some insight into the inversion process, consider first the "projection pursuit" approach (Kruskal, 1969) in which an unmixing vector is found that extracts the most non-Gaussian possible (maximally non-Gaussian) source signal. This source signal is then removed from the set of mixed signals and the process is repeated.

A way to implement this procedure is to consider the kurtosis ($K$) of the candidate source signal ($y = w^T u$) as a measure of non-Gaussianity

$$K = \frac{E[(y-\bar{y})^4]}{(E[(y-\bar{y})^2])^2} - 3 \qquad (7)$$

In this definition a Gaussian signal has $K = 0$. The process consists then of finding the unmixing vector which maximizes $|K|$ and thus finding the corresponding source signal. In contrast to the projection pursuit approach, ICA extracts $N$ source signals from $N$ mixed signals simultaneously.

The formal procedure decomposes the $n \times m$ data matrix $X(x,t) = [u_1 \cdots u_n]^T$ into

$$X(x,t) = A(t)S(x) \qquad (8)$$

where $A = [a_1, \cdots, a_k]$ is the $n \times k$ matrix of "mixing vectors" (MVs), which give the time evolution of the modes, and $k$ is the unknown number of independent modes. $S = [s_1, \cdots, s_k]^T$ is the $k \times m$ matrix of ICs that give the spatial structure of the modes (Funaro, Oja, and Valpola, 2003). Unlike PCA, ICA does not order the modes, and the amplitude of each IC and MV is defined only up to a multiplicative constant.

Before implementing the matrix inversion we must fix the number ($k$) of unknown independent components to be sought. This number can be determined by first performing a PCA and finding a break point in the eigenvalue spectrum. If there is no clear breakpoint in the spectrum we can keep the number of leading components that carry some arbitrary percentage of the variance (Hyvärinen, Karhunen, and Oja, 2001). With this information it is possible to define a $k \times n$ "whitening" matrix $V = D^{-1/2} E^T$, where the diagonal matrix $D$ has the $k$ leading eigenvalues of $\Lambda$ and the matrix $E = [e_1, \cdots e_k]$ has the corresponding $n$ dimensional PCs. The data can then be rotated to the $k \times m$ dimensional matrix $Z = VX = [z_1, \cdots, z_k]^T$ in which the whitened $m$-dimensional $z_i$ vectors have unit variance and are uncorrelated. The original problem is re-written in the form

$$Z = VX = VAS = WS \qquad (9)$$

and the goal is to solve for the $k \times k$ orthogonal matrix $W$, which permits then the calculation of the ICs and MVs in matrices $S$ and $A$, respectively. The matrix inversion is based on the FastICA method which maximizes the non-Gaussianity of the solutions by a fixed-point algorithm. The mathematical details of the algorithm are beyond the scope or interest of this paper (Hyvärinen, Karhunen, and Oja, 2001).

3.2 Application to Search for Active Longitudes

*3.2.1 Data Preparation and PCA*

In this analysis we make use of the WSO set of Carrington maps of photospheric magnetic fields. These are the maps that underlie the source-surface maps discussed in Section 2.2.2, and they come in the same format. Because the photospheric images are structurally complicated we do not analyze two-dimensional images directly. First, the 396 Carrington maps used (CR 1642 – 2037) were strung together end-to-end to make 30 latitudinal time series some 30 years in length with a 9.092 hour cadence.

We considered the absolute value of the field and used subsets of these data composed of averages of rasters 10 – 12, giving a single time series spanning latitudes +11.5° to +23.6° and averages of rasters 19 – 21 giving another time series spanning latitudes -11.5° to -23.6°. The limited latitude coverage is intended to further increase resolution in longitude/time. These particular rasters were selected because they span the latitudes that contain most of the late solar cycle fields. High-latitude features early in a given cycle will be missed.

Because we wish to look for rotation periods, if any, that might extend over more than one activity cycle, we divided all values in the two time series by a 27-day running average. This puts any signals during quiet-Sun times on an equal footing with those during active times.

The next step was to compute frequency spectra of the two time series. Because there are some missing data, we used a Lomb-Scargle periodogram program. For the northern hemisphere data, the program yields three periods with signal strengths above the 99.9% significance level. The strongest peak is at a period of 26.4 days, the second strongest is at 27.05 days, and the third is at 26.7 days. For the southern hemisphere data, interestingly, the strongest peak is at a long 28.2-day period. This is consistent with the result of Hoeksema and Scherrer (1987) that the southern hemisphere field rotated more slowly than the northern hemisphere during cycle 21. Other significant peaks are at periods shorter than 27 days. Because there has been interest in the literature (Neugebauer, *et al.*, 2000; Ruzmaikin *et al.*, 2001; Henney and Harvey, 2002) in a periodicity of 27.03 days seen in the solar wind, we chose to pursue that value. From this point we consider only the northern hemisphere data.

Because 27.03 days corresponds to 71.3 pixels in our time series, we cannot reproduce this periodicity. We clipped the series into the closest integer number of 71 pixel rasters. This corresponds to a 26.90-day periodicity which will be the focus of our study. The 406 rasters can be stacked sequentially, as in Figure 8a, to look for persistent features. Then, using each of the 71 pixel rasters of Figure 8a as an input vector, we carried out a PCA of the data. Notice that in this case the number of spatial pixels ($m = 71$) is less than the number of observations ($n = 406$). While the data are still centered in the spatial direction, we now obtain a spectrum of 71 eigenvalues. Figure 9 shows that break points occur at modes 4 and 6. The actual forms of the corresponding EOFs and PCs were not especially instructive (see Figure 11). We found that a five-mode reconstruction of the data, as in Equation (1), gives the result shown in Figure 8b in which the salient features of the original rasters are clearly displayed. These first five modes carry 32% of the variance.

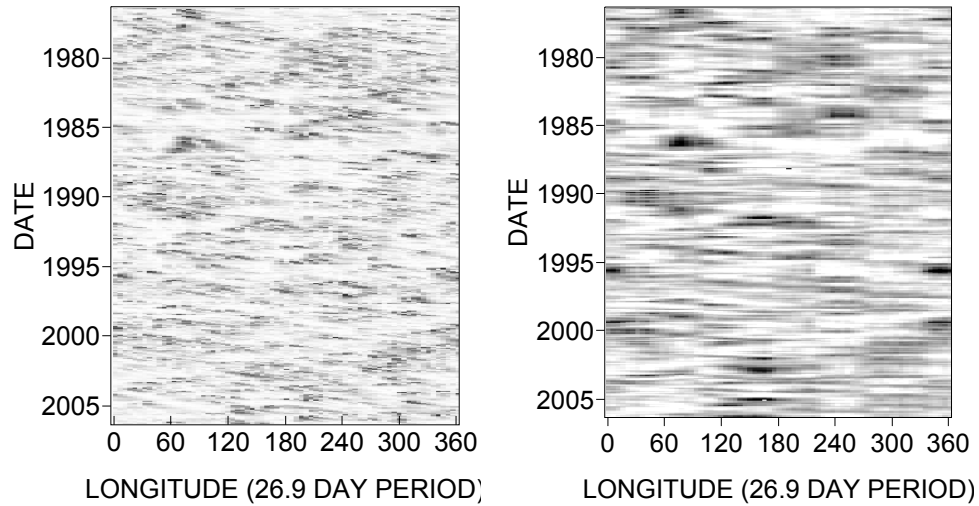

*Figure 8. a) Dark pixels indicate the presence of magnetic field in the solar latitude band from 11.5° to 23.6° North. The 71-pixel rasters correspond to a period of 26.90 days and span the period from 1976 to 2006. b) Five-mode reconstruction of the data.*

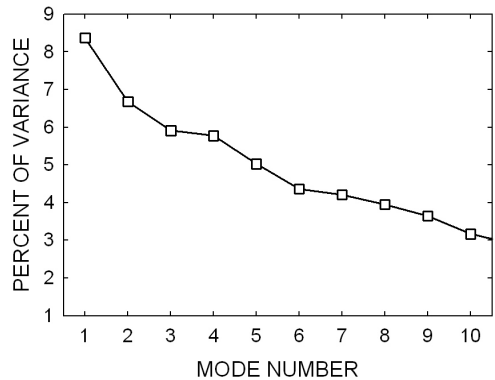

*Figure 9. Eigenvalue spectrum for the northern hemisphere latitudinal average of the WSO photospheric field. The eigenvalues give the percent of total variance explained by each mode.*

Because we have used 71 pixel rasters, a feature rotating around the Sun in 26.90 days would show as a vertical streak in Figure 8. Slower-rotating features with longer periods would slope down and to the right. A feature rotating with period 27.03 days would cross from left to right in 15.24 years. A feature rotating at the Carrington rate of 27.275 days would cross in 5.23 years. Although it might not be visible in Figure 8a, such rotating features can be revealed by the PCA as the presence of a pair of phased modes with close eigenvalues, as for modes 3 and 4 in Figure 9. We have carried out an analysis of the phase drift of this pair of modes, analogous to the analyses leading to Figure 7. We find that the phase drift corresponds to abrupt changes from one rotation period to another at intervals of about a year or two. The overall drift from the first rotation to the last gives an average rotation rate of 26.55 days, near the highest peak in the Lomb periodogram as described above.

To check this result we have applied a complex demodulation to this time series (*e.g.*, Bloomfield, 1976). This technique tests a signal by calculating its phase changes relative to a sine wave of selected constant frequency or period. To match our 71-pixel rasterization we choose the test period to be 26.90 days. The phase plot is presented in Figure 10. A constant phase indicates a period of 26.90 days; other nearby periods give different slopes; the overall slope from first to last phase gives period 26.5 days (the best fit from 1985 to 2005 gives 26.2 days). We thus find rough agreement on periods among the Lomb periodogram main peak (26.4 days), the PC phase advance (26.55 days) and the complex demodulation (26.5 days).

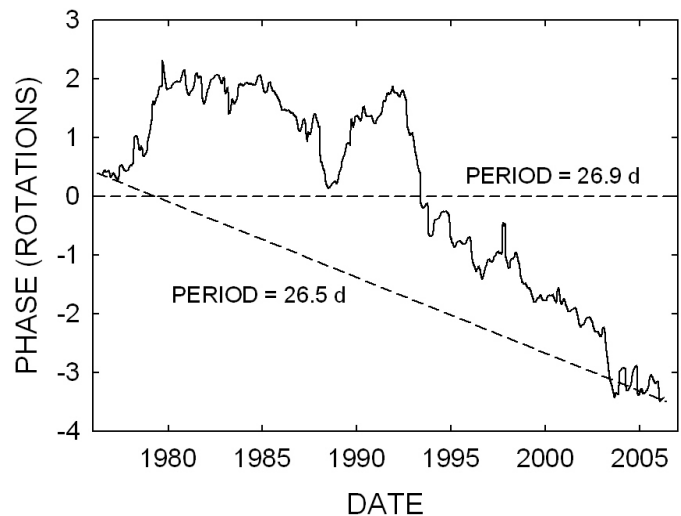

*Figure 10. Phase drift (solid line) in rotations versus date for the 11.5° to 23.6° North latitude band time series.*

*3.2.2 Independent Components*

We have seen in the preceding section that, for our purposes, the WSO longitudinal data can be adequately represented by keeping only the first five such modes. We thus achieve the first purpose of PCA, namely data reduction, acquiring a compact way to summarize the key data. However, the eigenvectors themselves, shown in Figure 11a, do not lend themselves to any particular interpretation. To attempt an interpretation, we apply ICA to the data by selecting the first four modes suggested by the first break point at mode 4 in the eigenvalue spectrum (Figure 9).

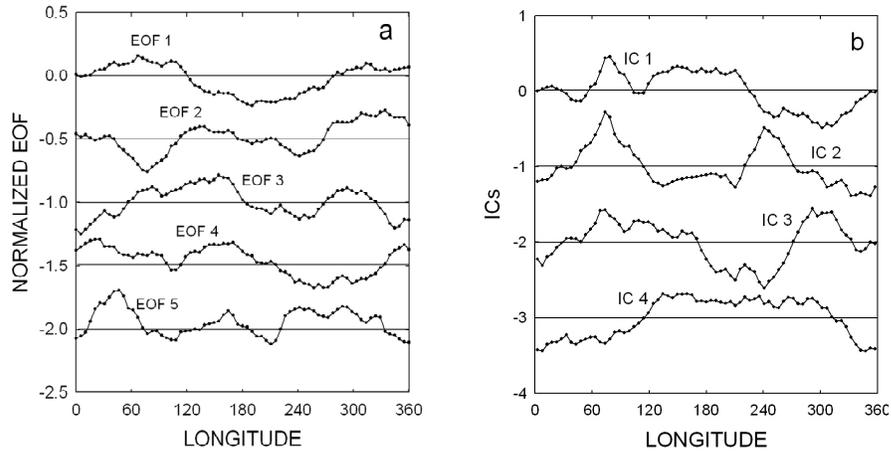

*Figure 11. (a) The first five EOFs for the PCA of the WSO +11.5° to +23.6° latitude average of WSO photospheric Carrington rotation maps, absolute values of line-of-sight fields. EOFs 2 - 5 have been shifted downward for clarity. (b) The ICs corresponding to four modes. ICs 2 – 4 have been shifted for clarity.*

The time dependences or MVs of the independent components are shown in Figure 12. The kurtosis of these four maximally independent, non-Gaussian sources are: -0.91, -0.81, -0.73, -0.28, so the first three are clearly the most non-Gaussian. The negative values indicate that the pdfs corresponding to the signals are "platykurtic," that is, more rectangular than Gaussian with very short tails. Positive values of kurtosis would indicate "leptokurtic" pdfs, with strong central peak and long tails. Note the general limitation $-2 \leq K < \infty$.

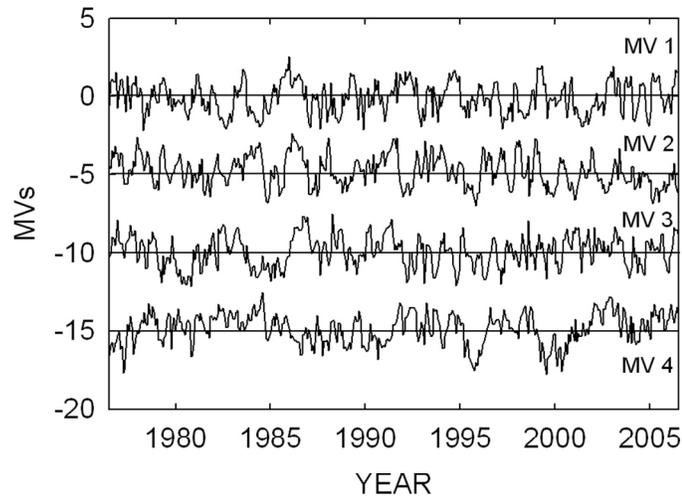

*Figure 12. The normalized time dependences of the four ICs in Figure 11b. MVs 2 through 4 have been shifted down for clarity.*

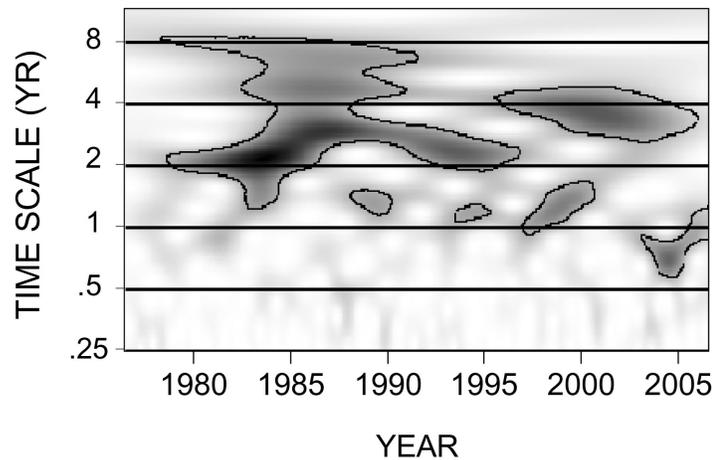

*Figure 13. Morlet wavelet power spectrum of MV 1 in Figure 12. The time is shown horizontally in years; the time scale, or period, of activity is plotted vertically. The wavelet power at times within about one time scale of the end points of the data is not reliable.*

To interpret the plots in Figure 12 we employ wavelet power spectra. Details are given in the widely used description by Torrence and Compo (1998). A Morlet wavelet power spectrum of MV 1 is shown in Figure 13. Areas of stronger wavelet power are shown as darker colors on a plot of time horizontally and time scale vertically. The dark contours enclose areas of 95% significance with respect to AR(1) red noise with coefficient $\alpha = 0.67$. The wavelet power at times within about one time scale of the end points of the data is not reliable.

The greatest spectral power for MV 1 occurs in the interval 1980 – 1985 with characteristic time scale of about two years. The wavelet power spectrum for MV 2 (not shown) shows significant power in the interval 1985 – 1990 at a scale of about two years, and that for MV 3 (also not shown) shows significant power between 1980 and 1990 at timescale about four years. The first three ICs all indicate some activity at longitude ≈ 75° with differing patterns otherwise. To the extent that ICA has truly isolated physically independent processes, all seem to be most active early in the data string and to refer to longitude near 75°. This corresponds to magnetic features seen in Figure 8. Note that IC 2 indicates two active longitudes (≈ 75° and ≈ 240°) on roughly opposite sides of the Sun, as is sometimes reported.

## 4. Summary and Conclusions

To examine global scale patterns of solar activity, we have described the main properties of Principal Components Analysis and Independent Component Analysis and have demonstrated applications of each to solar and solar – terrestrial physics. In particular we have used PCA to uncover related rotating modes in solar chromospheric magnetic fields and in the IMF polarity near one AU. During the active phase of solar cycle 23 the rotations of these two modes track one another in impressive detail.

To look for evidence of preferred longitudes of solar activity, we created a time series from the WSO synoptic Carrington maps of photospheric magnetic field. In doing this we narrowed our focus to the absolute values of the fields averaged over the latitudes from +11.5° to +23.6°. The time series was clipped into 71 pixel segments spanning 26.90 days. PCA of these segments indicated that the salient features of the data could be reproduced with just the five leading modes. Further, examination of two related modes (3 and 4) gave evidence of a rotation of the full pattern with a period around 26.5 days. This result was reinforced by a Lomb periodogram analysis and by complex demodulation.

Whereas PCA looks for global patterns correlated with the data, ICA searches for independent sources of the solar field. The first three of four ICs associated with the first four PCs emphasize particular longitudes. The time variations of the ICs indicate bursts of activity lasting from two to five years.

The PCA and ICA analyses we have described here examine sequences of individual quantities or images. Any set of analogous data could be used in a similar way. For example, synoptic maps of Fe XIV coronal green line emission or of coronal holes, and their modes, can be related to SW/IMF fluctuations.

We have treated the IMF data as vectors of 27 longitudes in a sequence of Bartels rotations. Instead, each "vector" could include simultaneous measurements of an array of different SW/IMF parameters, for example, a SW-density component, a temperature component, an IMF magnitude component, and so on. The vector would describe the state of the SW/IMF as measured at intervals of, say, one hour by ACE or other spacecraft. Modes of such vectors can be connected to changes in geomagnetic or solar-activity indices. Time ordering of the data vectors is not necessary either. For example, Fourier spectra of solar magnetograms can be used as input vectors that are ordered, not by time, but by total magnetic flux in the image. This illuminates how the size distributions of magnetic features differ in different regimes of total flux.

**Acknowledgements**

We are pleased to acknowledge and thank those who provided the data used in this work. These included the NASA OMNIWEB data center and the *Wind* and ACE spacecraft teams. We also made extensive use of photospheric and source-surface synoptic Carrington maps from the Wilcox Solar Observatory at Stanford University. We also thank the Laboratory of Computer and Information Science at Helsinki University of Technology, which offers the free FastICA software package in MATLAB (http://www.cis.hut.fi/projects/ica/fastica/).